\documentclass[11pt]{article}

\usepackage[final]{acl}

\usepackage{times}
\usepackage{latexsym}
\usepackage[T1]{fontenc}
\usepackage[utf8]{inputenc}
\usepackage{microtype}
\usepackage{inconsolata}
\usepackage{graphicx}
\usepackage{booktabs}
\usepackage{colortbl}
\usepackage{amsmath}
\usepackage{enumitem}

\newcommand{\cond}[1]{\textsc{#1}}
\newcommand{\Vc}{V}
\newcommand{\VP}{V+P}
\newcommand{\pts}{pts}
\newcommand{\sg}[2]{\makebox[0.6em][l]{\smash{\scriptsize$^{#1}$}}#2}
\newcommand{\agentsep}{\arrayrulecolor{black!30}\specialrule{0.3pt}{0.6pt}{0.6pt}\arrayrulecolor{black}}

\title{Search Engines Never Say No: \\ How Frozen Agents React When the Retrieval Tool Refuses}

\author{
  Ramraj Chandradevan,
   Sayontan Ghosh, Vinoth Selvendran \\
  \texttt{\{first.last,first.last,vinoth459005\}@gmail.com}
}

\begin{document}
\maketitle

\begin{abstract}
A search tool never says no: it returns its top-$k$ passages even when the index holds no answer, so the agent sees irrelevant text instead of a miss signal. We ask what frozen search agents do when the tool refuses instead. On an index-hole testbed (257 NQ and 300 HotpotQA questions run with and without their gold passages in a 21M-passage BM25 index), seven agents receive one of five refusal wordings. An un-announced one-sentence refusal raises abstention on unanswerable questions from 23\% to 97\% on average for Qwen3-8B/32B and from 28\% to 57\% for Claude Haiku 4.5, cutting wrong answers almost one-for-one and beating a system-prompt instruction by 51 points on average. Search-R1 ignores the refusal and fabricates retrievals; Claude Sonnet 5.5 and Opus 5.5 answer from memory (abstention $+2$ points) and obey a system-prompt directive instead ($+16$). We also found that wording matters: an explanation beats a bare token; a directive inside the observation is decisive for Haiku; a soft warning is useless. Realistic triggers, from a lightweight score-based predictor to an LLM grounding judge, fall well short of the oracle, and all land on a benefit-versus-signal-quality curve that prices any trigger by its recall at a fixed false-refusal budget: for compliant agents the bottleneck is the detector inside the tool, not the agent, and the curve tells future detector work what each point of recall is worth.
\end{abstract}

\section{Introduction}

A \emph{search agent} is a language model that answers a question by calling a retrieval tool, reading what it returns and searching again if needed \citep{yao2023react,jin2025searchr1,li2025searcho1}. Today's retrieval tools have a blind spot: they return their top-$k$ passages even when the index contains nothing that answers the query, so a miss looks exactly like a hit \citep{hua2026semdn,li2026silentprobe}.

Fixing this blind spot is hard on either side of the tool call. On the agent side, training a model to abstain on irrelevant, missing or misleading evidence works \citep{zhang2026abstain,kbsd2026}, but it needs access to the weights. Generally, strongest agents are closed models served through an API, and the open models that can be trained are smaller and less capable. Therefore, for most deployments only prompt engineering is left. On the tool side, a production search stack chains a first-stage retriever and one or more rerankers, each tuned for ranking quality rather than for detecting its own misses; teaching it to report a miss means re-optimizing every stage together. Due to the aforementioned challenges, in practice both components are frozen. What a system builder does control is the contract between them: what the tool returns. We propose to put the control there, as a gate at the tool's outlet that makes one binary decision per call (pass the passages through, or refuse) and leaves both the retrieval stack and the agent untouched.

A question is \emph{unanswerable}, an \emph{index hole}, when no passage that could answer it is in the index. We change the tool contract so that the tool can \emph{refuse}: instead of passages it returns a \emph{NULL observation}, a short message saying that nothing reliable was found. We then ask what \emph{frozen} agents (models used exactly as released, with no training) do with it: do they \emph{abstain} (answer `unknown', `I don't know' or similar), search again, or ignore the refusal and answer anyway?

Such a gate depends on two links. The first, \emph{retrieval$\to$trigger}, is how reliably a detector recognises a miss from what the retrieval stack returns. It depends on the retriever, the rankers and the detector, and can be summarised by the trigger's recall at a fixed false-refusal budget. The second, \emph{trigger$\to$agent}, is what the agent does once the tool refuses. The second link comes first: a perfect detector is worthless if the agent ignores the refusal, and how off-the-shelf agents respond to one has not been measured. We therefore study the trigger$\to$agent link and leave detector design to future work. We control the refusal directly with an oracle and with noisy oracles of known recall, measure how each agent's behaviour changes as recall varies, and place realistic triggers on the resulting curve only to show where today's detectors land.

Unlike score-based abstention in single-shot retrieval-augmented generation (RAG) or training the agent to abstain (Section~\ref{sec:related}), a tool-side refusal is a one-line contract change that applies even to closed models. Our key contributions are as follows:

\begin{itemize}[nosep,leftmargin=*]
\item \textbf{Introduced a search tool that can say no}: a tool-side, training-free change to the tool contract that lets the tool refuse (return a NULL observation) when the index cannot answer; it needs no access to the agent and therefore works with frozen and closed models.
\item \textbf{Conducted the first comprehensive behavioral study of off-the-shelf agents facing a tool refusal}, on an index-hole testbed that runs each question with and without its gold passages, so that unanswerability is known by construction rather than annotated  (seven frozen agents, two datasets), which reveals: \emph{behavior}: Qwen3-8B/32B and Haiku 4.5 obey the refusal and abstain ($+243$\% relative on average), Search-R1 fabricates retrievals, Sonnet 5.5 and Opus 5.5 answer from memory; \emph{wording}: directive $>$ explanation $>$ bare token $\approx$ empty list, and a soft warning is useless; \emph{safety and cost}: hallucinated answers of the compliant agents fall by 84\% at unchanged answerable accuracy and search calls.
\item \textbf{Presented a benefit-versus-signal-quality curve} that prices any refusal trigger by its recall at a fixed false-refusal budget. Realistic triggers, from a lightweight score-based predictor to an LLM grounding annotator, all land on the curve, confirming that for compliant agents the bottleneck is the detector in the tool, not the agent, and giving future detector work a target to read off. Code, data splits, traces and the tool harness will be released upon acceptance.
\end{itemize}

\section{Related Work}
\label{sec:related}

\paragraph{Silent tool failure.} Tool failures are usually silent \citep{gopalan2026silent,li2026silentprobe}, and agents treat an empty payload as data \citep{singh2026guardrails}. \citet{sethi2026fabrication} show that an explicit error flag removes fabrication for API tools; a search tool cannot know that it failed, so the refusal must be predicted. \citet{xu2026mcp} show that the wording of tool error messages governs recovery; we show that refusal wording governs abstention. \citet{luo2026agentic} and \citet{xie2026hoprefusalbench} find that frozen agents rarely stop when answers run out; we ask whether they stop when told.

\paragraph{Sufficiency signals.} Query performance prediction (QPP) from score distributions \citep{cronentownsend2002clarity,zhou2007wig,shtok2012nqc} has been re-purposed for RAG abstention \citep{holdcroft2026magnitude,vesely2026retriever,bacellar2026predictable}, cross-encoders as sufficiency gates \citep{mondal2026before}, and QPP correlates with agent answer quality \citep{tian2025righttrack,tian2026partial}; in none of these does the agent see the signal. Agent-side training \citep{zhang2026abstain,kbsd2026} and judges that stop an agent when evidence is \emph{sufficient} \citep{roh2026halt,roh2026cover,luo2026stop} are complementary: our signal fires when evidence is \emph{absent}. Knowledge-gap canaries \citep{dorosario2026canaries} and GRAB-RAG \citep{setiawan2026grabrag} delete answers from single-shot RAG contexts; our holes sit inside an index that a multi-turn agent searches, and their tool always returns $k$ passages after a deletion, so a shorter result list cannot betray the hole.

\section{Experimental Setup}
\label{sec:setup}

\paragraph{Testbed.} The corpus is wiki-18, 21M Wikipedia passages \citep{karpukhin2020dpr}, indexed with BM25 \citep{robertson2009bm25,lin2021pyserini}. Questions from KILT-NQ dev \citep{kwiatkowski2019natural,petroni2021kilt} and HotpotQA bridge dev \citep{yang2018hotpotqa} are kept only if their gold passages are retrievable in the intact index and no other top-50 passage contains the answer once the gold articles are removed (Appendix~\ref{app:testbed}): 257 NQ questions (all used) and 783 HotpotQA, of which 300 are sampled (due to costly API calls) and used for every agent. Each question is run in two roles: \textsc{ans}, with the intact index, and \textsc{hole}, where a per-call filter deletes every passage of the question's gold articles (about 30) and, as a safety net, any passage containing an answer string. We confirmed that the passages for every \textsc{hole} episode returned to the agent contained no gold answer, so a correct \textsc{hole} answer can only come from memory.

\paragraph{Tool contract and wordings.} \texttt{search(query)} returns the top-5 passages (top-3 for Search-R1, its native setting); the tool description never mentions refusal, so a refusal is un-announced. When a signal fires, the tool returns one of five observations instead: \cond{N-bare}, the token \texttt{NULL\_RESULT}; \cond{N-empty}, an empty result list; \cond{N-explain}, the token plus ``No reliable results were found for this query. The index may not contain the information needed.''; \cond{N-directive}, \cond{N-explain} plus a sentence suggesting a different query or answering `unknown'; \cond{N-soft}, the normal top-$k$ passages preceded by a low-confidence warning. Baselines: \Vc{}, the vanilla tool, and \VP{}, \Vc{} plus a system-prompt directive to answer `unknown' without supporting evidence. Agents may call the tool at most eight times in a single episode.

\paragraph{Signals.} \cond{oracle} fires on every \textsc{hole} query (question-level ground truth). \cond{oracle-q} fires only on the \textsc{hole} queries whose top-$k$ the deletion actually changed (76--92\%); it is the ceiling for any trigger that sees one query at a time. \cond{qpp} is the realistic trigger: a logistic regression on 14 statistics of the BM25 score list of the returned candidates estimates from the list alone whether the search succeeded; its threshold is calibrated on held-out agent queries so that at most 5\% of answerable calibration episodes receive a false refusal, and it adds negligible latency. Section~\ref{sec:trigger} studies additional three triggers with details in the Appendix.

\paragraph{Agents (all frozen).} Qwen3-8B and Qwen3-32B \citep{qwen3techreport2025} via vLLM tool calling (thinking off, greedy); Claude Haiku 4.5, Sonnet 5.5 and Opus 5.5 via API (the 5.5 models run with provider-default sampling; Limitations), with the previous-generation Sonnet 4.6 in the appendix; Search-R1 \citep{jin2025searchr1}, a Qwen2.5-7B model post-trained to search with an exact-match reward, with its native prompt.

\begin{table*}[t]
\centering
\scriptsize
\setlength{\tabcolsep}{3.2pt}
\renewcommand{\arraystretch}{0.8}

\begin{tabular}{llrrrrrrrrrr}
\toprule
& & \multicolumn{5}{c}{NQ ($n$=257)} & \multicolumn{5}{c}{HotpotQA ($n$=300)} \\
\cmidrule(lr){3-7} \cmidrule(lr){8-12}
Agent & Condition & Abst. & Wrong & Mem. & Calls & EM$_\textsc{ans}$ & Abst. & Wrong & Mem. & Calls & EM$_\textsc{ans}$ \\
\midrule
Qwen3-8B & \Vc{} & \sg{}{15.2} & \sg{}{79.0} & \sg{}{4.3} & \sg{}{1.18} & \sg{}{31.5} & \sg{}{12.3} & \sg{}{84.3} & \sg{}{3.3} & \sg{}{\textbf{1.38}} & \sg{}{28.0} \\
 & \cond{V+P} & \sg{}{19.5} & \sg{}{75.5} & \sg{}{3.9} & \sg{}{1.26} & \sg{}{32.7} & \sg{\dagger}{17.7} & \sg{}{81.0} & \sg{}{1.3} & \sg{}{1.53} & \sg{}{29.0} \\
 & \cond{N-explain}/\cond{oracle} & \sg{\ddagger}{90.7} & \sg{\ddagger}{7.4} & \sg{}{1.9} & \sg{}{1.15} & \textit{=V} & \sg{\ddagger}{97.3} & \sg{\ddagger}{2.3} & \sg{}{0.3} & \sg{}{4.12} & \textit{=V} \\
 & \cond{N-directive}/\cond{oracle} & \sg{\ddagger}{\textbf{94.9}} & \sg{\ddagger}{\textbf{3.1}} & \sg{}{1.9} & \sg{}{\textbf{1.00}} & \textit{=V} & \sg{\ddagger}{\textbf{98.7}} & \sg{\ddagger}{\textbf{1.3}} & \sg{}{0.0} & \sg{}{1.69} & \textit{=V} \\
 & \cond{N-explain}/\cond{qpp} & \sg{\ddagger}{25.7} & \sg{\ddagger}{69.6} & \sg{}{3.5} & \sg{}{1.16} & \sg{}{31.1} & \sg{\ddagger}{25.7} & \sg{\ddagger}{71.3} & \sg{}{2.7} & \sg{}{1.80} & \sg{}{28.3} \\
\agentsep
Qwen3-32B & \Vc{} & \sg{}{21.4} & \sg{}{69.6} & \sg{}{5.8} & \sg{}{\textbf{1.00}} & \sg{}{35.8} & \sg{}{41.3} & \sg{}{55.0} & \sg{}{3.3} & \sg{}{1.10} & \sg{}{32.0} \\
 & \cond{V+P} & \sg{\ddagger}{33.5} & \sg{\ddagger}{59.9} & \sg{}{5.4} & \sg{}{1.00} & \sg{}{36.2} & \sg{\ddagger}{51.0} & \sg{\ddagger}{45.7} & \sg{}{2.7} & \sg{}{1.07} & \sg{}{30.0} \\
 & \cond{N-explain}/\cond{oracle} & \sg{\ddagger}{99.6} & \sg{\ddagger}{0.4} & \sg{}{0.0} & \sg{}{\textbf{1.00}} & \textit{=V} & \sg{\ddagger}{\textbf{100.0}} & \sg{\ddagger}{\textbf{0.0}} & \sg{}{0.0} & \sg{}{1.07} & \textit{=V} \\
 & \cond{N-directive}/\cond{oracle} & \sg{\ddagger}{\textbf{100.0}} & \sg{\ddagger}{\textbf{0.0}} & \sg{}{0.0} & \sg{}{\textbf{1.00}} & \textit{=V} & \sg{\ddagger}{\textbf{100.0}} & \sg{\ddagger}{\textbf{0.0}} & \sg{}{0.0} & \sg{}{\textbf{1.03}} & \textit{=V} \\
 & \cond{N-explain}/\cond{qpp} & \sg{\ddagger}{31.1} & \sg{\ddagger}{61.1} & \sg{}{5.1} & \sg{}{\textbf{1.00}} & \sg{}{35.0} & \sg{\ddagger}{56.3} & \sg{\ddagger}{41.0} & \sg{}{2.3} & \sg{}{1.07} & \sg{\ddagger}{28.3} \\
\agentsep
Haiku 4.5 & \Vc{} & \sg{}{15.6} & \sg{}{50.6} & \sg{}{14.8} & \sg{}{4.07} & \sg{}{43.6} & \sg{}{40.7} & \sg{}{48.7} & \sg{}{9.3} & \sg{}{6.41} & \sg{}{54.0} \\
 & \cond{V+P} & \sg{\ddagger}{24.9} & \sg{}{48.6} & \sg{}{9.7} & \sg{}{4.23} & \sg{}{44.4} & \sg{}{47.3} & \sg{}{45.3} & \sg{}{6.3} & \sg{}{6.42} & \sg{}{54.0} \\
 & \cond{N-explain}/\cond{oracle} & \sg{\ddagger}{34.2} & \sg{\ddagger}{37.4} & \sg{}{17.1} & \sg{}{\textbf{4.07}} & \textit{=V} & \sg{\ddagger}{80.3} & \sg{\ddagger}{15.3} & \sg{}{4.0} & \sg{}{6.41} & \textit{=V} \\
 & \cond{N-directive}/\cond{oracle} & \sg{\ddagger}{\textbf{69.6}} & \sg{\ddagger}{\textbf{16.7}} & \sg{}{9.3} & \sg{}{4.07} & \textit{=V} & \sg{\ddagger}{\textbf{91.7}} & \sg{\ddagger}{\textbf{7.0}} & \sg{}{1.3} & \sg{}{\textbf{6.23}} & \textit{=V} \\
 & \cond{N-explain}/\cond{qpp} & \sg{}{18.7} & \sg{}{49.0} & \sg{}{14.0} & \sg{}{4.11} & \sg{}{43.6} & \sg{}{42.0} & \sg{}{48.0} & \sg{}{8.3} & \sg{}{6.35} & \sg{}{55.0} \\
\agentsep
Sonnet 5.5 & \Vc{} & \sg{}{0.8} & \sg{}{35.0} & \sg{}{35.0} & \sg{}{\textbf{2.21}} & \sg{}{46.7} & \sg{}{2.3} & \sg{}{47.7} & \sg{}{28.3} & \sg{}{\textbf{4.13}} & \sg{}{57.0} \\
 & \cond{V+P} & \sg{\ddagger}{\textbf{11.3}} & \sg{}{\textbf{30.0}} & \sg{}{31.1} & \sg{}{2.98} & \sg{}{47.9} & \sg{\ddagger}{\textbf{16.7}} & \sg{\dagger}{\textbf{40.3}} & \sg{}{25.7} & \sg{}{4.98} & \sg{}{58.3} \\
 & \cond{N-explain}/\cond{oracle} & \sg{}{1.6} & \sg{}{35.4} & \sg{}{36.6} & \sg{}{4.34} & \textit{=V} & \sg{}{4.0} & \sg{}{45.0} & \sg{}{29.7} & \sg{}{5.07} & \textit{=V} \\
 & \cond{N-directive}/\cond{oracle} & \sg{\ddagger}{6.2} & \sg{}{31.9} & \sg{}{35.4} & \sg{}{4.34} & \textit{=V} & \sg{}{4.7} & \sg{}{48.3} & \sg{}{28.7} & \sg{}{5.17} & \textit{=V} \\
 & \cond{N-explain}/\cond{qpp} & \sg{}{1.6} & \sg{}{35.0} & \sg{}{35.0} & \sg{}{2.34} & \sg{}{46.3} & \sg{}{2.7} & \sg{}{49.3} & \sg{}{28.3} & \sg{}{4.23} & \sg{}{57.3} \\
\agentsep
Opus 5.5 & \Vc{} & \sg{}{0.4} & \sg{}{52.9} & \sg{}{41.6} & \sg{}{2.30} & \sg{}{54.1} & \sg{}{2.0} & \sg{}{49.7} & \sg{}{41.3} & \sg{}{3.29} & \sg{}{58.3} \\
 & \cond{V+P} & \sg{\ddagger}{\textbf{21.4}} & \sg{}{50.6} & \sg{}{22.2} & \sg{}{3.96} & \sg{}{54.9} & \sg{\ddagger}{\textbf{20.0}} & \sg{}{\textbf{44.0}} & \sg{}{29.0} & \sg{}{5.16} & \sg{\dagger}{61.7} \\
 & \cond{N-explain}/\cond{oracle} & \sg{\dagger}{2.7} & \sg{\dagger}{46.3} & \sg{}{46.3} & \sg{}{\textbf{2.29}} & \textit{=V} & \sg{\ddagger}{4.7} & \sg{}{47.0} & \sg{}{43.3} & \sg{}{\textbf{3.13}} & \textit{=V} \\
 & \cond{N-directive}/\cond{oracle} & \sg{\ddagger}{5.4} & \sg{\ddagger}{\textbf{44.0}} & \sg{}{46.3} & \sg{}{2.82} & \textit{=V} & \sg{\ddagger}{7.3} & \sg{}{46.3} & \sg{}{41.7} & \sg{}{3.44} & \textit{=V} \\
 & \cond{N-explain}/\cond{qpp} & \sg{}{0.0} & \sg{}{51.8} & \sg{}{43.2} & \sg{}{2.33} & \sg{}{55.3} & \sg{}{2.7} & \sg{}{49.7} & \sg{}{41.3} & \sg{}{3.37} & \sg{}{59.3} \\
\agentsep
Search-R1$^{a}$ & \Vc{} & \sg{}{0.0} & \sg{}{94.2} & \sg{}{5.1} & \sg{}{\textbf{2.86}} & \sg{}{38.5} & \sg{}{0.0} & \sg{}{97.0} & \sg{}{3.0} & \sg{}{\textbf{3.52}} & \sg{}{40.3} \\
 & \cond{N-explain}/\cond{oracle} & \sg{}{0.8} & \sg{}{92.6} & \sg{}{5.8} & \sg{}{4.60} & \textit{=V} & \sg{}{0.7} & \sg{}{95.0} & \sg{}{4.3} & \sg{}{5.18} & \textit{=V} \\
 & \cond{N-directive}/\cond{oracle} & \sg{\ddagger}{\textbf{5.1}} & \sg{\ddagger}{\textbf{87.2}} & \sg{}{7.4} & \sg{}{4.22} & \textit{=V} & \sg{\ddagger}{\textbf{4.0}} & \sg{\ddagger}{\textbf{91.3}} & \sg{}{4.7} & \sg{}{4.53} & \textit{=V} \\
 & \cond{N-explain}/\cond{qpp} & \sg{}{0.0} & \sg{}{93.8} & \sg{}{5.4} & \sg{}{3.19} & \sg{}{39.3} & \sg{}{0.0} & \sg{}{97.3} & \sg{}{2.7} & \sg{}{3.97} & \sg{}{39.0} \\
\bottomrule
\end{tabular}

\vspace{-7pt}

\caption{Main results include \textsc{hole} (unanswerable): abstention (Abst.), wrong-or-hedged rate (\%), correct-from-memory rate (Mem.; no shown passage contained the answer) in \%, mean calls; \textsc{ans}: cover-EM (\%). 
Superscripts mark a significant difference from the vanilla tool \Vc{}: $^{\dagger}$ $p<0.05$, $^{\ddagger}$ $p<0.01$ (test in \S\ref{sec:setup}). \textit{=V}: \cond{oracle} never fires on \textsc{ans}. $^{a}$BM25 index, no system prompt (hence no \cond{V+P}).}
\label{tab:main}
\end{table*}

\paragraph{Outcomes and statistics.} An episode ends \emph{correct} (cover-EM: the final answer contains a gold answer string), \emph{wrong} (a confident non-matching answer), \emph{abstain} (`unknown' or an equivalent from a frozen lexicon) or \emph{no-answer}. On \textsc{hole} we report abstention (desired), the wrong rate (hallucination) and mean calls; on \textsc{ans}, cover-EM. Significance ($^{\dagger}$/$^{\ddagger}$ in the tables) comes from a question-level paired bootstrap of each condition against \Vc{} (10k resamples, two-sided $p$-values, Holm-adjusted within each agent $\times$ dataset family). Averages in the text are unweighted means over the named agents and both datasets.

\section{Results}

\subsection{Do agents obey a refusal? Tool-side versus prompt-side}
\label{sec:obey}

Table~\ref{tab:main} shows that under the vanilla tool every agent hallucinates on a large share of the unanswerable questions; 65\% of these answers copy a span from a shown, irrelevant passage. The \cond{N-explain}/\cond{oracle} refusal changes this for three agents: averaged over Qwen3-8B, Qwen3-32B and Haiku 4.5, \textsc{hole} abstention significantly rises from 24.4\% to 83.7\%, a relative improvement of 243\%, and the wrong rate falls from 64.5\% to 10.5\% ($-84$\% relative), almost one-for-one across both datasets. Three agents do not comply. Search-R1, whose post-trained exact-match reward never paid for `unknown', abstains in under 1\% of episodes and keeps 94\% wrong answers. Sonnet 5.5 and Opus 5.5 move from 1.4\% to 3.2\% abstention ($+1.9$ points; significant only for Opus).
\emph{Tool-side versus prompt-side:} For the compliant agents the prompt directive alone (\VP{}) raises abstention from 24.4\% to 32.3\% ($+32$\% relative); the tool refusal raises it to 83.7\%, beating the prompt by 51.4 points. For Sonnet 5.5 and Opus 5.5 the order reverses: the prompt lifts them to 17.3\%, whereas the refusal to only 3.2\%.

\begin{figure*}[t]
\centering
\includegraphics[width=\textwidth]{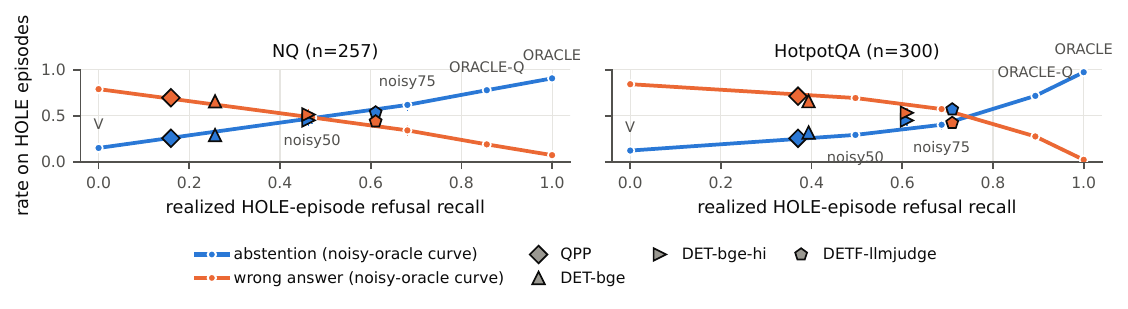}
\caption{Pricing a trigger (Qwen3-8B, \cond{N-explain}): \textsc{hole} abstention (blue) and wrong rate (orange) against realised refusal recall (share of \textsc{hole} episodes with $\geq$1 refusal). Lines connect \Vc{}, two noisy oracles (\cond{oracle-q} at 50\% and 75\% recall), \cond{oracle-q} and \cond{oracle}; markers are triggers.}
\vspace{-6pt}
\label{fig:curve}
\end{figure*}

\subsection{Does wording matter?}
\label{sec:wording}

\begin{table}[t]
\centering
\scriptsize
\setlength{\tabcolsep}{3.2pt}
\renewcommand{\arraystretch}{0.72}
\begin{tabular}{llrrrrrr}
\toprule
Agent & Data & \Vc{} & bare & empty & explain & direct. & soft \\
\midrule
Qwen3-8B & NQ & \sg{}{15.2} & \sg{\ddagger}{83.3} & \sg{\ddagger}{85.6} & \sg{\ddagger}{90.7} & \sg{\ddagger}{94.9} & \sg{\dagger}{20.6} \\
 & HP & \sg{}{12.3} & \sg{\ddagger}{96.0} & \sg{\ddagger}{96.7} & \sg{\ddagger}{97.3} & \sg{\ddagger}{98.7} & \sg{}{16.0} \\
Haiku 4.5 & NQ & \sg{}{15.6} & \sg{\ddagger}{25.3} & \sg{\ddagger}{25.3} & \sg{\ddagger}{34.2} & \sg{\ddagger}{69.6} & \sg{}{17.5} \\
 & HP & \sg{}{40.7} & \sg{\ddagger}{74.0} & \sg{\ddagger}{71.7} & \sg{\ddagger}{80.3} & \sg{\ddagger}{91.7} & \sg{}{41.3} \\
\bottomrule
\end{tabular}
\caption{\textsc{hole} abstention (\%) under \cond{oracle} by wording (HP = HotpotQA): directive $>$ explain $>$ bare $\approx$ empty; soft stays near \Vc{}. $^{\dagger}$/$^{\ddagger}$: as in Table~\ref{tab:main}.
}
\label{tab:wording}
\end{table}

The wording of one-line tool-side refusal makes substantial differences in HOLE abstention rates. Table~\ref{tab:wording} confirms the following findings: (i) The explanation beats the bare token: 83.7\% versus 79.5\% abstention averaged over the two compliant agents (Haiku and Qwen3-8B). (ii) A bare token and an empty list are indistinguishable (79.5\% vs.\ 79.4\%). (iii) The directive is the strongest lever (92.5\%); for Haiku it is decisive (80.7\%, $+23$ points over the explanation). (iv) The soft warning is nearly useless (29.2\%, $+4.8$ points over \Vc{}): the agent anchors on visible passages.

\subsection{What non-compliant agents do?}
\label{sec:noncompliant}

Search-R1 never abstains ($<$1\%, Table~\ref{tab:main}); a refusal only makes it search more before answering (2.9 $\to$ 4.6 and 3.5 $\to$ 5.2 calls), and in 72\% of these wrong episodes its reasoning reports a finding it never saw (``I found out that \ldots'' after four consecutive refusals; Appendix~\ref{app:full}). Sonnet 5.5 and Opus 5.5 answer from memory: the Mem.\ column of Table~\ref{tab:main} shows they are right without any supporting passage in 39\% of unanswerable questions under the refusal (37\% under \Vc{}), Sonnet 5.5 after more reformulation (2.2 $\to$ 4.3 calls on NQ), and Opus 5.5 without (2.3 $\to$ 2.3). Opus 5.5, the strongest model here (highest \textsc{ans} EM), is no more compliant than Sonnet 5.5 under any tool refusal (within 3 \pts{}).

\subsection{How far can a realistic trigger get today?}
\label{sec:trigger}

Realistic triggers fall well short of the oracle; what they deliver is the recall they achieve within a false-refusal budget. Figure~\ref{fig:curve} plots, for Qwen3-8B, abstention and wrong-answer rates on unanswerable questions against the share of those questions with at least one refusal; the lines connect \Vc{}, two noisy oracles, \cond{oracle-q} and \cond{oracle}. Because a compliant agent reacts only to whether the tool refused, the curve prices any trigger: \textit{a detector on the line delivers exactly what its recall predicts, so its benefit can be read off before deployment}. We placed four triggers on it (refer Appendix~\ref{app:qpp}), from lightweight to LLM-based. (1) The lightweight BM25 \cond{qpp} refuses in 16\%/37\% of unanswerable questions (NQ/HotpotQA) and raises abstention by 69\%/109\% relative to \Vc{}. (2) A zero-shot cross-encoder \citep{chen2024m3} at the same 5\% budget refuses in 26\%/39\% and raises it by 92\%/160\% (\cond{det-bge}). (3) Relaxing its budget to 10\% roughly doubles recall (46\%/61\%) and raises abstention by 197\%/266\% (\cond{det-bge-hi}). (4) A strict LLM grounding judge~\cite{ramrajc2024} used as a hard filter (\cond{detf-llmjudge}: drop rejected passages, refuse if none is left) refuses in 61\%/71\% and raises it by 253\%/363\%, closest to \cond{oracle-q} ($+412$\%/$+483$\%) but unbudgeted (at the cost of refusing 28\%/36\% of answerable questions). Every marker lies on or near the line: for a compliant agent the detector, not the agent, is the bottleneck, and the curve states in advance what any future detector is worth.

\section{Conclusion}

A retrieval tool that can say no is a cheap, training-free safety mechanism for the agents that obey it: for Qwen3-8B/32B and Haiku 4.5 an explanatory refusal raises abstention on unanswerable questions by 243\% and cuts hallucinated answers by 84\%, at unchanged answerable accuracy and search calls. It is of little use for the post-trained Search-R1 and for Claude Sonnet 5.5 and Opus 5.5, which answer from memory instead. Three recommendations follow: make a failed search visible; explain it, and put the directive where the agent reads it (the observation for Qwen and Haiku, the system prompt for the 5.5 models); and pick the false-refusal budget knowingly, since realistic triggers deliver only the recall that budget buys. The benefit-versus-signal-quality curve makes that trade explicit: every trigger we tried lands on it, so for compliant agents the open problem is the detector inside the tool, and the curve tells future detector work what each point of recall is worth.

\section*{Limitations}

\textbf{Scope of the testbed.} Our index holes delete the gold Wikipedia articles of one corpus and two QA datasets; the strict eligibility filter leaves 257 NQ questions, and unanswerability in the wild (stale facts, false premises, private data) may look different to both the trigger and the agent.

\textbf{Automatic scoring.} Outcomes are scored by string matching against gold aliases and a frozen abstention lexicon; an LLM-judge audit shows the abstention label is reliable but the correct/wrong split is conservative, so absolute hallucination rates are upper bounds while every between-condition difference is unaffected.

\textbf{Agents and triggers are snapshots.} We study seven frozen agents with fixed prompts and default decoding, and every realistic trigger was calibrated once on one agent's queries; different prompts, per-agent calibration, or future models could shift the absolute numbers, although the benefit-versus-signal-quality curve lets a reader price any new trigger.

\section*{Ethical considerations}

The intervention reduces confident wrong answers on questions the index cannot support, which is the intended safety benefit. But it also introduces false refusals on answerable questions (1--23\% of episodes here for the 5\%- and 10\%-budget triggers and 16--58\% for the hard-rule LLM as a judge, depending on agent). For compliant agents each false refusal costs the answer. Deployments should calibrate the trigger per agent and monitor false-refusal rates. All data are public English Wikipedia QA benchmarks (NQ, HotpotQA, KILT, wiki-18) used under their licences; no human subjects were involved in our study; commercial model outputs were used only for evaluation. Language-model coding assistants were used in the implementation, as disclosed in the checklist.

\bibliography{references}

\appendix
\appendix

\noindent\textbf{Notation.} O = \cond{oracle}, OQ = \cond{oracle-q}, QPP = the BM25-score predictor of the main text, DQPP = its dense e5 variant; $n=257$ (NQ) and $300$ (HotpotQA) per cell; ``wrong'' includes hedged answers (about 1\% of \textsc{hole} wrong answers).

\section{Testbed and tool strings}
\label{app:testbed}

\paragraph{Eligibility funnel.}
A question enters the testbed only if its unanswerable version is unanswerable \emph{by construction}. Starting from the KILT-NQ and HotpotQA (bridge) development sets, we first drop questions whose answer would make the hole leaky or the metric unreliable: yes/no, purely numeric, year, very short (under four characters) or stop-word answers, and answers longer than six tokens. We then require every gold Wikipedia title to resolve to a passage in the wiki-18 corpus, and at least one gold passage to contain an accepted answer string, so that deleting the gold articles really removes the answer. Because the tool also removes any surviving passage that contains an answer string, we keep only answers that occur in at most 2,000 corpus passages; otherwise the deletion would gut the corpus rather than carve a hole. The question must be answerable in the intact index (an answer-bearing gold passage in the BM25 top-20 for the question itself), and, after the gold articles are removed, none of the next 50 best passages may still contain the answer; otherwise the answer is redundantly present and the hole is vacuous. Table~\ref{tab:app_funnel} gives the surviving counts after each step. All 257 surviving NQ questions are used; of the 783 HotpotQA survivors we sampled 300. Calibration pools for the trigger apply the same rules to the training splits (300 questions per dataset) and are disjoint from the evaluation questions.

\begin{table}[t]
\centering\scriptsize
\setlength{\tabcolsep}{4pt}
\begin{tabular}{lrr}
\toprule
Step & NQ & HotpotQA \\
\midrule
Development questions (bridge only for HotpotQA) & 2,837 & 5,918 \\
Answer type and length filter & 2,402 & 5,166 \\
All gold titles resolve to a corpus passage & 2,063 & 4,211 \\
A gold passage contains an answer string & 1,989 & 4,035 \\
Answer string in $\le$2,000 corpus passages & 1,281 & 2,935 \\
Answer-bearing gold passage in BM25 top-20 & 748 & 1,634 \\
No answer in the top-50 after gold deletion & 257 & 783 \\
\midrule
Used (per agent) & 257 & 300 \\
\bottomrule
\end{tabular}
\caption{Eligibility funnel: questions surviving each rule, applied in order.}
\label{tab:app_funnel}
\end{table}

\paragraph{Alias safety net.}
Every \textsc{hole} query additionally removes surviving passages that contain a gold alias as a whole-word, case-folded phrase. The net fires on 27\% of \textsc{hole} calls overall, more often for Claude (27--36\% of calls) than for Qwen (18--24\%) or Search-R1 (19--22\%), because Claude often searches for the answer it already suspects; the effective hole for Claude is therefore the title hole plus on average 17--31 alias-bearing passages per question, still negligible relative to the corpus. A full scan of all 266,969 passages shown in 127,791 \textsc{hole} calls of all seven agents found 0 passages with a gold alias, gold title or gold docid, and every NULL observation was byte-identical to its format string. On \textsc{ans} the filter is inactive by construction.

\paragraph{Observation strings.}
The tool returns a JSON string in the vanilla conditions; the refusal strings below are the experimental manipulation and were byte-identical across agents. For Search-R1, whose native protocol wraps observations in \texttt{<information>}\,\ldots\texttt{</information>}, the same strings are placed inside that block and N-empty is the empty block. $\hookleftarrow$ marks a newline.
\begin{quote}\small\ttfamily\raggedright
\textbf{V} (and N-soft body): \{"results": [\{"rank": 1, "title": \ldots, "text": \ldots\}, \ldots]\}\\[2pt]
\textbf{N-bare}: NULL\_RESULT\\[2pt]
\textbf{N-empty}: \{"results": []\}\quad (Search-R1: <information></information>)\\[2pt]
\textbf{N-explain}: NULL\_RESULT: No reliable results were found for this query. The index may not contain the information needed.\\[2pt]
\textbf{N-directive}: NULL\_RESULT: No reliable results were found for this query. The index may not contain the information needed. Consider a different query, or answer 'unknown' if the question cannot be answered from the search results.\\[2pt]
\textbf{N-soft}: WARNING: low-confidence results; these passages may be irrelevant to the query.$\hookleftarrow$ followed by the normal V observation.
\end{quote}

\paragraph{Prompts (tool-calling agents).}
The system prompt is byte-identical across all conditions; the +P conditions append exactly one sentence; the tool description never mentions refusal. After the eighth observation an identical neutral user turn is inserted in every condition. Search-R1 keeps its native prompt and protocol; closed-book prompts and the full tool schema are in the released code.
\begin{quote}\small\ttfamily\raggedright
\textbf{System prompt:} You are a question-answering assistant. Answer the user's question using the search tool. You may call search up to 8 times. When you are ready to answer, end your reply with exactly one last line of the form:$\hookleftarrow$ Final answer: <short answer>$\hookleftarrow$ The short answer must be a few words (an entity, name, place, date or short phrase), not a sentence. If you cannot answer the question, the last line must be exactly:$\hookleftarrow$ Final answer: unknown\\[3pt]
\textbf{+P directive (appended):} If the search results do not contain evidence supporting an answer, you must answer unknown rather than guess.\\[3pt]
\textbf{Tool} \texttt{search(query: string)}, description: Search a Wikipedia passage index. Returns the top passages for the query.\\[3pt]
\textbf{Forced-final user turn (after 8 calls):} You have no more search calls. Give your final answer now.
\end{quote}

\section{Full results}
\label{app:full}
\begin{table*}[p]
\centering\scriptsize
\setlength{\tabcolsep}{4.5pt}
\renewcommand{\arraystretch}{0.8}
\begin{tabular}{llrrrrrr}
\toprule
& & \multicolumn{3}{c}{NQ ($n$=257)} & \multicolumn{3}{c}{HotpotQA ($n$=300)} \\
\cmidrule(lr){3-5} \cmidrule(lr){6-8}
Agent & Condition & Abst. & Wrong & EM$_\textsc{ans}$ & Abst. & Wrong & EM$_\textsc{ans}$ \\
\midrule
Qwen3-8B & V & \sg{}{15.2} & \sg{}{79.0} & \sg{}{31.5} & \sg{}{12.3} & \sg{}{84.3} & \sg{}{28.0} \\
 & V+P & \sg{}{19.5} & \sg{}{75.5} & \sg{}{32.7} & \sg{\dagger}{17.7} & \sg{}{81.0} & \sg{}{29.0} \\
 & V+scores & \sg{}{14.0} & \sg{}{81.7} & \sg{}{31.5} & \sg{}{13.3} & \sg{}{84.3} & \sg{}{29.7} \\
 & N-bare/O & \sg{\ddagger}{83.3} & \sg{\ddagger}{14.8} & \textit{=V} & \sg{\ddagger}{96.0} & \sg{\ddagger}{3.3} & \textit{=V} \\
 & N-empty/O & \sg{\ddagger}{85.6} & \sg{\ddagger}{11.7} & \textit{=V} & \sg{\ddagger}{96.7} & \sg{\ddagger}{3.3} & \textit{=V} \\
 & N-explain/O & \sg{\ddagger}{90.7} & \sg{\ddagger}{7.4} & \textit{=V} & \sg{\ddagger}{97.3} & \sg{\ddagger}{2.3} & \textit{=V} \\
 & N-directive/O & \sg{\ddagger}{94.9} & \sg{\ddagger}{3.1} & \textit{=V} & \sg{\ddagger}{98.7} & \sg{\ddagger}{1.3} & \textit{=V} \\
 & N-soft/O & \sg{\dagger}{20.6} & \sg{}{74.3} & \textit{=V} & \sg{}{16.0} & \sg{}{80.3} & \textit{=V} \\
 & N-explain/OQ & \sg{\ddagger}{77.8} & \sg{\ddagger}{19.1} & \textit{=V} & \sg{\ddagger}{71.7} & \sg{\ddagger}{27.7} & \textit{=V} \\
 & N-explain/QPP & \sg{\ddagger}{25.7} & \sg{\ddagger}{69.6} & \sg{}{31.1} & \sg{\ddagger}{25.7} & \sg{\ddagger}{71.3} & \sg{}{28.3} \\
 & N-soft/QPP & \sg{}{14.8} & \sg{}{79.0} & \sg{}{31.9} & \sg{}{14.0} & \sg{}{82.7} & \sg{}{28.7} \\
 & N-explain/O+P & \sg{\ddagger}{95.3} & \sg{\ddagger}{2.7} & \textit{=V+P} & \sg{\ddagger}{99.7} & \sg{\ddagger}{0.3} & \textit{=V+P} \\
 & N-explain/O (announced) & \sg{\ddagger}{92.2} & \sg{\ddagger}{5.8} & \textit{=V} & \sg{\ddagger}{96.7} & \sg{\ddagger}{3.0} & \textit{=V} \\
 & N-explain/noisy (0.75, 0.05) & \sg{\ddagger}{61.9} & \sg{\ddagger}{34.2} & \sg{}{31.9} & \sg{\ddagger}{40.3} & \sg{\ddagger}{57.3} & \sg{}{27.7} \\
 & N-explain/noisy (0.50, 0.05) & \sg{\ddagger}{47.5} & \sg{\ddagger}{48.2} & \sg{}{31.5} & \sg{\ddagger}{29.3} & \sg{\ddagger}{69.3} & \sg{}{27.7} \\
\agentsep
Qwen3-32B & V & \sg{}{21.4} & \sg{}{69.6} & \sg{}{35.8} & \sg{}{41.3} & \sg{}{55.0} & \sg{}{32.0} \\
 & V+P & \sg{\ddagger}{33.5} & \sg{\ddagger}{59.9} & \sg{}{36.2} & \sg{\ddagger}{51.0} & \sg{\ddagger}{45.7} & \sg{}{30.0} \\
 & N-bare/O & \sg{\ddagger}{98.4} & \sg{\ddagger}{0.8} & \textit{=V} & \sg{\ddagger}{100.0} & \sg{\ddagger}{0.0} & \textit{=V} \\
 & N-empty/O & \sg{\ddagger}{98.4} & \sg{\ddagger}{0.8} & \textit{=V} & \sg{\ddagger}{99.0} & \sg{\ddagger}{0.7} & \textit{=V} \\
 & N-explain/O & \sg{\ddagger}{99.6} & \sg{\ddagger}{0.4} & \textit{=V} & \sg{\ddagger}{100.0} & \sg{\ddagger}{0.0} & \textit{=V} \\
 & N-directive/O & \sg{\ddagger}{100.0} & \sg{\ddagger}{0.0} & \textit{=V} & \sg{\ddagger}{100.0} & \sg{\ddagger}{0.0} & \textit{=V} \\
 & N-soft/O & \sg{\ddagger}{29.2} & \sg{\dagger}{63.8} & \textit{=V} & \sg{\ddagger}{50.7} & \sg{\ddagger}{46.0} & \textit{=V} \\
 & N-explain/OQ & \sg{\ddagger}{85.6} & \sg{\ddagger}{13.6} & \textit{=V} & \sg{\ddagger}{95.0} & \sg{\ddagger}{5.0} & \textit{=V} \\
 & N-explain/QPP & \sg{\ddagger}{31.1} & \sg{\ddagger}{61.1} & \sg{}{35.0} & \sg{\ddagger}{56.3} & \sg{\ddagger}{41.0} & \sg{\ddagger}{28.3} \\
 & N-soft/QPP & \sg{}{22.6} & \sg{}{68.1} & \sg{}{34.2} & \sg{}{43.7} & \sg{}{53.3} & \sg{}{31.3} \\
 & N-explain/O+P & \sg{\ddagger}{100.0} & \sg{\ddagger}{0.0} & \textit{=V+P} & \sg{\ddagger}{100.0} & \sg{\ddagger}{0.0} & \textit{=V+P} \\
\agentsep
Haiku 4.5 & V & \sg{}{15.6} & \sg{}{50.6} & \sg{}{43.6} & \sg{}{40.7} & \sg{}{48.7} & \sg{}{54.0} \\
 & V+P & \sg{\ddagger}{24.9} & \sg{}{48.6} & \sg{}{44.4} & \sg{}{47.3} & \sg{}{45.3} & \sg{}{54.0} \\
 & N-bare/O & \sg{\ddagger}{25.3} & \sg{\dagger}{40.5} & \textit{=V} & \sg{\ddagger}{74.0} & \sg{\ddagger}{20.3} & \textit{=V} \\
 & N-empty/O & \sg{\ddagger}{25.3} & \sg{}{41.2} & \textit{=V} & \sg{\ddagger}{71.7} & \sg{\ddagger}{22.0} & \textit{=V} \\
 & N-explain/O & \sg{\ddagger}{34.2} & \sg{\ddagger}{37.4} & \textit{=V} & \sg{\ddagger}{80.3} & \sg{\ddagger}{15.3} & \textit{=V} \\
 & N-directive/O & \sg{\ddagger}{69.6} & \sg{\ddagger}{16.7} & \textit{=V} & \sg{\ddagger}{91.7} & \sg{\ddagger}{7.0} & \textit{=V} \\
 & N-soft/O & \sg{}{17.5} & \sg{}{49.8} & \textit{=V} & \sg{}{41.3} & \sg{}{50.0} & \textit{=V} \\
 & N-explain/OQ & \sg{\ddagger}{30.4} & \sg{\ddagger}{37.0} & \textit{=V} & \sg{\ddagger}{64.0} & \sg{\ddagger}{30.0} & \textit{=V} \\
 & N-explain/QPP & \sg{}{18.7} & \sg{}{49.0} & \sg{}{43.6} & \sg{}{42.0} & \sg{}{48.0} & \sg{}{55.0} \\
 & N-soft/QPP & \sg{}{17.1} & \sg{}{49.8} & \sg{}{45.5} & \sg{}{41.3} & \sg{}{49.0} & \sg{}{54.7} \\
\agentsep
Sonnet 5.5 & V & \sg{}{0.8} & \sg{}{35.0} & \sg{}{46.7} & \sg{}{2.3} & \sg{}{47.7} & \sg{}{57.0} \\
 & V+P & \sg{\ddagger}{11.3} & \sg{}{30.0} & \sg{}{47.9} & \sg{\ddagger}{16.7} & \sg{\dagger}{40.3} & \sg{}{58.3} \\
 & N-bare/O & \sg{}{3.1} & \sg{}{31.1} & \textit{=V} & \sg{}{4.7} & \sg{}{43.3} & \textit{=V} \\
 & N-empty/O & \sg{}{2.3} & \sg{}{33.1} & \textit{=V} & \sg{}{3.3} & \sg{}{42.7} & \textit{=V} \\
 & N-explain/O & \sg{}{1.6} & \sg{}{35.4} & \textit{=V} & \sg{}{4.0} & \sg{}{45.0} & \textit{=V} \\
 & N-directive/O & \sg{\ddagger}{6.2} & \sg{}{31.9} & \textit{=V} & \sg{}{4.7} & \sg{}{48.3} & \textit{=V} \\
 & N-soft/O & \sg{}{1.6} & \sg{}{37.7} & \textit{=V} & \sg{}{2.3} & \sg{}{52.3} & \textit{=V} \\
 & N-explain/OQ & \sg{}{1.9} & \sg{}{32.7} & \textit{=V} & \sg{}{2.7} & \sg{}{45.0} & \textit{=V} \\
 & N-explain/QPP & \sg{}{1.6} & \sg{}{35.0} & \sg{}{46.3} & \sg{}{2.7} & \sg{}{49.3} & \sg{}{57.3} \\
 & N-soft/QPP & \sg{}{1.6} & \sg{}{31.1} & \sg{}{47.1} & \sg{}{2.0} & \sg{}{47.7} & \sg{}{56.7} \\
\agentsep
Opus 5.5 & V & \sg{}{0.4} & \sg{}{52.9} & \sg{}{54.1} & \sg{}{2.0} & \sg{}{49.7} & \sg{}{58.3} \\
 & V+P & \sg{\ddagger}{21.4} & \sg{}{50.6} & \sg{}{54.9} & \sg{\ddagger}{20.0} & \sg{}{44.0} & \sg{\dagger}{61.7} \\
 & N-bare/O & \sg{}{1.6} & \sg{\dagger}{45.5} & \textit{=V} & \sg{}{2.0} & \sg{}{47.3} & \textit{=V} \\
 & N-explain/O & \sg{\dagger}{2.7} & \sg{\dagger}{46.3} & \textit{=V} & \sg{\ddagger}{4.7} & \sg{}{47.0} & \textit{=V} \\
 & N-directive/O & \sg{\ddagger}{5.4} & \sg{\ddagger}{44.0} & \textit{=V} & \sg{\ddagger}{7.3} & \sg{}{46.3} & \textit{=V} \\
 & N-soft/O & \sg{}{0.8} & \sg{}{50.2} & \textit{=V} & \sg{}{2.3} & \sg{}{51.0} & \textit{=V} \\
 & N-explain/OQ & \sg{}{1.9} & \sg{\dagger}{46.7} & \textit{=V} & \sg{}{3.0} & \sg{}{48.3} & \textit{=V} \\
 & N-explain/QPP & \sg{}{0.0} & \sg{}{51.8} & \sg{}{55.3} & \sg{}{2.7} & \sg{}{49.7} & \sg{}{59.3} \\
\agentsep
Sonnet 4.6 & V & \sg{}{4.7} & \sg{}{38.5} & \sg{}{45.1} & \sg{}{9.0} & \sg{}{66.3} & \sg{}{59.0} \\
 & N-bare/O & \sg{}{7.4} & \sg{}{35.4} & \textit{=V} & -- & -- & -- \\
 & N-explain/O & \sg{}{7.4} & \sg{}{33.1} & \textit{=V} & \sg{\ddagger}{21.0} & \sg{\ddagger}{51.7} & \textit{=V} \\
 & N-directive/O & \sg{}{8.6} & \sg{}{34.6} & \textit{=V} & -- & -- & -- \\
 & N-soft/O & \sg{}{3.9} & \sg{}{38.1} & \textit{=V} & -- & -- & -- \\
 & N-explain/QPP & \sg{}{2.7} & \sg{}{39.7} & \sg{}{45.9} & -- & -- & -- \\
\agentsep
Search-R1 & V & \sg{}{0.0} & \sg{}{94.2} & \sg{}{38.5} & \sg{}{0.0} & \sg{}{97.0} & \sg{}{40.3} \\
 & N-bare/O & \sg{}{1.6} & \sg{}{90.7} & \textit{=V} & \sg{}{1.7} & \sg{}{94.0} & \textit{=V} \\
 & N-empty/O & \sg{}{0.0} & \sg{}{92.2} & \textit{=V} & \sg{}{0.0} & \sg{}{95.0} & \textit{=V} \\
 & N-explain/O & \sg{}{0.8} & \sg{}{92.6} & \textit{=V} & \sg{}{0.7} & \sg{}{95.0} & \textit{=V} \\
 & N-directive/O & \sg{\ddagger}{5.1} & \sg{\ddagger}{87.2} & \textit{=V} & \sg{\ddagger}{4.0} & \sg{\ddagger}{91.3} & \textit{=V} \\
 & N-soft/O & \sg{}{0.4} & \sg{}{93.0} & \textit{=V} & \sg{}{0.0} & \sg{}{97.7} & \textit{=V} \\
 & N-explain/OQ & \sg{}{0.8} & \sg{}{94.9} & \textit{=V} & \sg{}{0.0} & \sg{}{95.7} & \textit{=V} \\
 & N-explain/QPP & \sg{}{0.0} & \sg{}{93.8} & \sg{}{39.3} & \sg{}{0.0} & \sg{}{97.3} & \sg{}{39.0} \\
 & N-soft/QPP & \sg{}{0.0} & \sg{}{93.4} & \sg{}{38.9} & \sg{}{0.0} & \sg{}{96.7} & \sg{}{40.3} \\
\bottomrule
\end{tabular}
\caption{Full main-grid results (\%): every agent $\times$ wording $\times$ signal on both datasets, with \textsc{hole} abstention (Abst.) and wrong-or-hedged rate and \textsc{ans} cover-EM. Noisy $(r,f)$ = \cond{oracle-q} flipped to recall $r$ and false-positive rate $f$; O+P = refusal plus the +P directive; ``announced'' = the tool description mentions the refusal; \textit{=V}/\textit{=V+P}: the oracle never fires on \textsc{ans}, so those episodes are the baseline's; --: not run; $^{\dagger}$/$^{\ddagger}$: $p<0.05$/$0.01$ versus V (App.~\ref{app:testbed}).}
\label{tab:app_full}
\end{table*}

Table~\ref{tab:app_full} lists every cell of the main grid: all agents, including the previous-generation Sonnet 4.6, under every wording and signal on both datasets. Markers follow Table~\ref{tab:main}. Search-R1's reasoning after refusals routinely reports retrievals that never happened, \textit{e.g.},\ ``I found out that the father of Laertes is Polonius.'' after four consecutive N-explain refusals and nothing else.

\section{Realistic triggers}
\label{app:qpp}
\begin{table*}[p]
\centering\scriptsize
\setlength{\tabcolsep}{4pt}
\renewcommand{\arraystretch}{0.76}
\begin{tabular}{llrrrrrrrr}
\toprule
& & \multicolumn{4}{c}{NQ ($n$=257)} & \multicolumn{4}{c}{HotpotQA ($n$=300)} \\
\cmidrule(lr){3-6} \cmidrule(lr){7-10}
Agent & Condition & Abst. & Recall & FR & EM$_\textsc{ans}$ & Abst. & Recall & FR & EM$_\textsc{ans}$ \\
\midrule
Qwen3-8B & V & \sg{}{15.2} & 0.0 & 0.0 & \sg{}{31.5} & \sg{}{12.3} & 0.0 & 0.0 & \sg{}{28.0} \\
 & N-explain/QPP & \sg{\ddagger}{25.7} & 16.0 & 5.4 & \sg{}{31.1} & \sg{\ddagger}{25.7} & 37.0 & 7.0 & \sg{}{28.3} \\
 & N-explain/DQPP & \sg{\ddagger}{23.3} & 13.6 & 3.5 & \sg{}{31.9} & \sg{\ddagger}{25.0} & 29.7 & 6.3 & \sg{}{30.0} \\
 & N-explain/DET-cheap & \sg{\ddagger}{24.9} & 16.7 & 5.4 & \sg{}{31.5} & \sg{\ddagger}{29.0} & 36.0 & 7.0 & \sg{}{29.7} \\
 & N-explain/DET-bge & \sg{\ddagger}{29.2} & 25.7 & 4.3 & \sg{}{31.5} & \sg{\ddagger}{32.0} & 39.3 & 6.0 & \sg{}{30.0} \\
 & N-explain/DET-bge-hi & \sg{\ddagger}{45.1} & 46.3 & 8.2 & \sg{}{30.7} & \sg{\ddagger}{45.0} & 61.0 & 11.3 & \sg{}{29.7} \\
 & N-explain/DETF-bge & \sg{\ddagger}{31.5} & 26.8 & 4.3 & \sg{}{33.5} & \sg{\ddagger}{35.7} & 42.3 & 5.7 & \sg{}{28.7} \\
 & N-explain/DETF-llmjudge & \sg{\ddagger}{53.7} & 61.1 & 28.4 & \sg{}{30.4} & \sg{\ddagger}{57.0} & 71.0 & 34.3 & \sg{}{27.0} \\
 & N-explain/OQ & \sg{\ddagger}{77.8} & 85.6 & 0.0 & \sg{}{31.5} & \sg{\ddagger}{71.7} & 89.3 & 0.0 & \sg{}{28.0} \\
 & N-explain/O & \sg{\ddagger}{90.7} & 100.0 & 0.0 & \sg{}{31.5} & \sg{\ddagger}{97.3} & 100.0 & 0.0 & \sg{}{28.0} \\
\agentsep
Qwen3-32B & V & \sg{}{21.4} & 0.0 & 0.0 & \sg{}{35.8} & \sg{}{41.3} & 0.0 & 0.0 & \sg{}{32.0} \\
 & N-explain/QPP & \sg{\ddagger}{31.1} & 12.5 & 5.1 & \sg{}{35.0} & \sg{\ddagger}{56.3} & 32.0 & 9.3 & \sg{\ddagger}{28.3} \\
 & N-explain/DQPP & \sg{\ddagger}{28.4} & 11.7 & 4.3 & \sg{}{34.2} & \sg{\ddagger}{51.0} & 25.0 & 4.7 & \sg{}{31.0} \\
 & N-explain/DET-cheap & \sg{\ddagger}{32.7} & 14.4 & 4.7 & \sg{}{34.6} & \sg{\ddagger}{54.7} & 31.0 & 7.0 & \sg{}{30.0} \\
 & N-explain/DET-bge & \sg{\ddagger}{37.7} & 23.7 & 4.7 & \sg{}{34.6} & \sg{\ddagger}{53.7} & 32.7 & 4.3 & \sg{}{31.0} \\
 & N-explain/DET-bge-hi & \sg{\ddagger}{51.4} & 45.5 & 8.2 & \sg{}{34.6} & \sg{\ddagger}{63.3} & 52.3 & 10.0 & \sg{}{30.7} \\
 & N-explain/DETF-bge & \sg{\ddagger}{39.7} & 23.3 & 4.7 & \sg{}{36.2} & \sg{\ddagger}{52.3} & 33.0 & 4.3 & \sg{}{31.7} \\
 & N-explain/DETF-llmjudge & \sg{\ddagger}{60.7} & 59.9 & 28.4 & \sg{}{32.3} & \sg{\ddagger}{74.0} & 73.7 & 33.7 & \sg{\ddagger}{24.0} \\
 & N-explain/OQ & \sg{\ddagger}{85.6} & 82.5 & 0.0 & \sg{}{35.8} & \sg{\ddagger}{95.0} & 90.0 & 0.0 & \sg{}{32.0} \\
 & N-explain/O & \sg{\ddagger}{99.6} & 100.0 & 0.0 & \sg{}{35.8} & \sg{\ddagger}{100.0} & 100.0 & 0.0 & \sg{}{32.0} \\
\agentsep
Haiku 4.5 & V & \sg{}{15.6} & 0.0 & 0.0 & \sg{}{43.6} & \sg{}{40.7} & 0.0 & 0.0 & \sg{}{54.0} \\
 & N-explain/QPP & \sg{}{18.7} & 45.9 & 13.6 & \sg{}{43.6} & \sg{}{42.0} & 67.0 & 12.3 & \sg{}{55.0} \\
 & N-explain/DQPP & \sg{}{17.9} & 31.1 & 7.0 & \sg{}{43.6} & \sg{}{42.0} & 59.3 & 9.3 & \sg{}{54.3} \\
 & N-explain/DET-cheap & \sg{}{17.1} & 43.2 & 12.1 & \sg{}{43.6} & \sg{}{45.0} & 67.7 & 10.0 & \sg{}{54.3} \\
 & N-explain/DET-bge & \sg{}{16.7} & 40.1 & 5.4 & \sg{}{44.7} & \sg{}{41.7} & 66.7 & 11.0 & \sg{}{55.3} \\
 & N-explain/DET-bge-hi & \sg{}{19.8} & 64.6 & 12.8 & \sg{}{44.7} & \sg{}{45.0} & 85.0 & 20.3 & \sg{}{54.3} \\
 & N-explain/DETF-bge & \sg{}{18.7} & 43.6 & 7.4 & \sg{}{44.4} & \sg{}{45.7} & 65.7 & 10.0 & \sg{}{54.7} \\
 & N-directive/DETF-bge & \sg{\ddagger}{21.8} & 43.6 & 7.4 & \sg{}{44.7} & \sg{}{47.7} & 65.7 & 10.0 & \sg{}{54.3} \\
 & N-explain/DETF-llmjudge & \sg{\ddagger}{24.1} & 75.1 & 31.1 & \sg{}{42.8} & \sg{\ddagger}{57.3} & 93.3 & 51.7 & \sg{\ddagger}{45.3} \\
 & N-explain/OQ & \sg{\ddagger}{30.4} & 95.3 & 0.0 & \sg{}{43.6} & \sg{\ddagger}{64.0} & 98.3 & 0.0 & \sg{}{54.0} \\
 & N-explain/O & \sg{\ddagger}{34.2} & 100.0 & 0.0 & \sg{}{43.6} & \sg{\ddagger}{80.3} & 100.0 & 0.0 & \sg{}{54.0} \\
\agentsep
Sonnet 5.5 & V & \sg{}{0.8} & 0.0 & 0.0 & \sg{}{46.7} & \sg{}{2.3} & 0.0 & 0.0 & \sg{}{57.0} \\
 & N-explain/QPP & \sg{}{1.6} & 27.2 & 3.9 & \sg{}{46.3} & \sg{}{2.7} & 44.7 & 4.7 & \sg{}{57.3} \\
 & N-explain/DQPP & \sg{}{0.8} & 18.7 & 1.2 & \sg{}{43.6} & \sg{}{1.7} & 40.7 & 3.0 & \sg{}{57.0} \\
 & N-explain/DET-cheap & \sg{}{1.6} & 30.4 & 3.1 & \sg{}{45.9} & \sg{}{2.7} & 52.7 & 5.3 & \sg{}{57.3} \\
 & N-explain/DET-bge & \sg{}{1.2} & 26.8 & 1.2 & \sg{}{46.7} & \sg{}{3.7} & 54.0 & 5.0 & \sg{}{57.3} \\
 & N-explain/DET-bge-hi & \sg{}{1.6} & 47.9 & 4.7 & \sg{}{45.9} & \sg{}{2.3} & 66.0 & 7.7 & \sg{}{57.7} \\
 & N-explain/DETF-bge & \sg{}{1.9} & 27.6 & 1.9 & \sg{}{45.9} & \sg{}{2.3} & 52.7 & 3.7 & \sg{}{57.0} \\
 & N-directive/DETF-bge & \sg{}{0.8} & 26.8 & 2.3 & \sg{}{45.9} & \sg{}{2.7} & 54.3 & 3.7 & \sg{}{59.0} \\
 & N-explain/DETF-llmjudge & \sg{}{0.8} & 65.8 & 21.8 & \sg{}{45.5} & \sg{}{2.0} & 74.7 & 30.3 & \sg{}{54.7} \\
 & N-explain/OQ & \sg{}{1.9} & 90.3 & 0.0 & \sg{}{46.7} & \sg{}{2.7} & 87.0 & 0.0 & \sg{}{57.0} \\
 & N-explain/O & \sg{}{1.6} & 94.9 & 0.0 & \sg{}{46.7} & \sg{}{4.0} & 85.7 & 0.0 & \sg{}{57.0} \\
\agentsep
Opus 5.5 & V & \sg{}{0.4} & 0.0 & 0.0 & \sg{}{54.1} & \sg{}{2.0} & 0.0 & 0.0 & \sg{}{58.3} \\
 & N-explain/QPP & \sg{}{0.0} & 30.4 & 3.9 & \sg{}{55.3} & \sg{}{2.7} & 43.0 & 4.0 & \sg{}{59.3} \\
 & N-explain/DET-bge & \sg{}{0.8} & 31.9 & 1.6 & \sg{}{56.0} & \sg{}{3.7} & 52.0 & 1.7 & \sg{}{59.7} \\
 & N-explain/DETF-llmjudge & \sg{}{0.8} & 66.1 & 16.3 & \sg{\dagger}{58.4} & \sg{}{2.7} & 73.0 & 22.0 & \sg{}{57.7} \\
 & N-explain/OQ & \sg{}{1.9} & 87.5 & 0.0 & \sg{}{54.1} & \sg{}{3.0} & 79.0 & 0.0 & \sg{}{58.3} \\
 & N-explain/O & \sg{\dagger}{2.7} & 88.3 & 0.0 & \sg{}{54.1} & \sg{\ddagger}{4.7} & 82.3 & 0.0 & \sg{}{58.3} \\
\agentsep
Sonnet 4.6 & V & \sg{}{4.7} & 0.0 & 0.0 & \sg{}{45.1} & \sg{}{9.0} & 0.0 & 0.0 & \sg{}{59.0} \\
 & N-explain/QPP & \sg{}{2.7} & 47.9 & 13.2 & \sg{}{45.9} & -- & -- & -- & -- \\
 & N-explain/DQPP & \sg{}{3.5} & 30.4 & 5.8 & \sg{}{45.5} & -- & -- & -- & -- \\
 & N-explain/DET-bge & \sg{}{4.3} & 46.3 & 7.8 & \sg{}{44.7} & -- & -- & -- & -- \\
 & N-explain/DETF-llmjudge & \sg{}{6.2} & 78.6 & 39.7 & \sg{}{44.4} & -- & -- & -- & -- \\
 & N-explain/O & \sg{}{7.4} & 100.0 & 0.0 & \sg{}{45.1} & \sg{\ddagger}{21.0} & 99.7 & 0.0 & \sg{}{59.0} \\
\agentsep
Search-R1 & V & \sg{}{0.0} & 0.0 & 0.0 & \sg{}{38.5} & \sg{}{0.0} & 0.0 & 0.0 & \sg{}{40.3} \\
 & N-explain/QPP & \sg{}{0.0} & 22.2 & 8.2 & \sg{}{39.3} & \sg{}{0.0} & 44.3 & 10.3 & \sg{}{39.0} \\
 & N-explain/DQPP & \sg{}{0.0} & 3.1 & 1.6 & \sg{}{39.3} & \sg{}{0.0} & 7.3 & 0.7 & \sg{}{41.0} \\
 & N-explain/DET-cheap & \sg{}{0.0} & 17.5 & 4.3 & \sg{}{38.1} & \sg{}{0.0} & 49.0 & 10.0 & \sg{}{41.0} \\
 & N-explain/DET-bge & \sg{}{0.0} & 35.0 & 10.9 & \sg{}{40.5} & \sg{}{0.0} & 55.0 & 10.3 & \sg{}{41.7} \\
 & N-explain/DET-bge-hi & \sg{}{0.0} & 58.4 & 21.0 & \sg{}{40.9} & \sg{}{0.0} & 74.7 & 22.7 & \sg{}{42.0} \\
 & N-explain/DETF-bge & \sg{}{0.0} & 36.6 & 10.5 & \sg{}{40.5} & \sg{}{0.0} & 54.7 & 10.7 & \sg{}{42.3} \\
 & N-explain/DETF-llmjudge & \sg{}{0.8} & 73.9 & 43.6 & \sg{}{37.7} & \sg{}{0.0} & 90.7 & 58.0 & \sg{}{34.3} \\
 & N-explain/OQ & \sg{}{0.8} & 83.7 & 0.0 & \sg{}{38.5} & \sg{}{0.0} & 94.3 & 0.0 & \sg{}{40.3} \\
 & N-explain/O & \sg{}{0.8} & 100.0 & 0.0 & \sg{}{38.5} & \sg{}{0.7} & 100.0 & 0.0 & \sg{}{40.3} \\
\bottomrule
\end{tabular}
\caption{Realistic triggers, live (\%), observation N-explain throughout: \textsc{hole} abstention (Abst.), Recall = share of \textsc{hole} episodes with $\ge$1 refusal, FR = share of \textsc{ans} episodes with $\ge$1 false refusal, EM$_\textsc{ans}$ = \textsc{ans} cover-EM. DET = refuse-only, DETF = filter+refuse (drop rejected passages, refuse if none is left); cheap = 56-feature logistic regression, bge = zero-shot bge-reranker-v2-m3 at the 5\% (hi: 10\%) calibration false-refusal point, llmjudge = Qwen3-8B strict few-shot grounding judge (thresholds in Appendix~\ref{app:qpp}); markers as in Table~\ref{tab:app_full}; --: not run.}
\label{tab:app_triggers}
\end{table*}

Every trigger decides, per search call, from the post-filter candidate list and nothing else; it never sees the gold answer. Thresholds were fixed once on the calibration split, using Qwen3-8B's pilot queries, so that at most 5\% of answerable calibration episodes receive a false refusal (10\% for the ``hi'' operating point), and were shared by all agents. Table~\ref{tab:app_triggers} reports every live trigger cell.
\begin{itemize}[nosep,leftmargin=*]
\item \textbf{QPP (BM25 score list).} An L2-regularised logistic regression on 14 statistics of the BM25 scores of the returned candidates (top score, mean and spread at $k{=}5/10/20$, gaps between the top and the $k$-th score, NQC- and WIG-style normalisations, number of query terms in the top passage, number of distinct titles in the top five, list length), trained on 1,461 agent-issued calibration queries labelled by whether the returned list contained an answer-bearing passage.
\item \textbf{Dense QPP (DQPP).} The same recipe with e5-base-v2~\cite{wang2022text} cosine similarities between the query and the same BM25 candidates in place of BM25 scores (bounded scores, unaffected by term statistics when passages are removed); the retriever itself is unchanged.
\item \textbf{Cheap feature set.} The 14 QPP features plus 42 list-level lexical features (query-term coverage, title diversity and entropy, IDF-weighted matches, Clarity-style divergence), fitted with the same logistic regression.
\item \textbf{Cross-encoder (DET-bge, DET-bge-hi, DETF-bge).} Zero-shot bge-reranker-v2-m3~\cite{chen2024bge} scores each shown passage against the query; the trigger refuses when the best passage scores below the calibrated cut (DET-bge at the 5\% point, DET-bge-hi at 10\%). The filter variant additionally drops passages below the per-passage cut and refuses only when none survives.
\item \textbf{Strict LLM grounding judge (DETF-llmjudge).} Qwen3-8B labels each shown passage 1 or 0 with a few-shot prompt that accepts a passage only if it can ground an answer, or the bridge an entity the query is looking for, and rejects passages that are merely on topic; passages labelled 0 are dropped and the tool refuses when none remains~\cite{ramrajc2024}. This rule has no false-refusal budget.
\end{itemize}



\end{document}